\documentclass[twocolumn]{article}
\usepackage{mathptmx}
\usepackage{url}
\usepackage{graphicx}
\usepackage{color}
\usepackage{fixltx2e}
\usepackage{array}
\newcommand{\bq}{\begin{equation}}
\newcommand{\eq}{\end{equation}}

\newcommand{\GBS}{GBytes/sec}

\newcommand{\GHZ}{GHz}

\newcommand{\GB}{GB}
\newcommand{\KB}{kB}
\newcommand{\MB}{MB}

\usepackage{listings}
\begin{document}
\title{Optimizing ccNUMA locality for task-parallel execution under OpenMP and TBB on multicore-based systems}
\author{Markus Wittmann and Georg Hager\\Erlangen Regional Computing Center, 91058 Erlangen, Germany}
\date{August 2, 2010}

\maketitle

\begin{abstract}
Task parallelism as employed by the OpenMP task
construct or some Intel Threading Building Blocks (TBB) components, although
ideal for tackling
irregular problems or typical producer/consumer schemes, bears some
potential for performance bottlenecks if locality of data access is
important, which is typically the case for memory-bound code on
ccNUMA systems.  We present a thin software layer
ameliorates adverse effects of dynamic task distribution by sorting
tasks into locality queues, each of which is preferably processed by
threads that belong to the same locality domain. Dynamic scheduling
is fully preserved inside each domain, and is preferred over
possible load imbalance even if nonlocal access is required, making
this strategy well-suited for typical multicore-mutisocket systems. The
effectiveness of the approach is demonstrated by using a blocked
six-point stencil solver as a toy model. 
\end{abstract}
%
\section{Introduction}

\subsection{Dynamic scheduling on ccNUMA systems}

``Cache-coherent nonuniform memory access'' (ccNUMA) is the preferred
system architecture for multisocket shared-memory servers today.  In
ccNUMA, main memory is logically shared, meaning that all memory
locations can be accessed by all sockets and cores in the system
transparently. However, since main memory is physically distributed,
i.e., partitioned in so-called \emph{locality domains} (LDs), access
bandwidths and latencies may vary, depending on which core accesses a
certain part of memory. Access is fastest from the cores directly
attached to a domain. Nonlocal accesses are mediated by some
inter-domain network, which is also capable of maintaining
cache coherency throughout the system.

The big advantage of ccNUMA is that the available main memory
bandwidth scales with the number of LDs, and shared-memory nodes with
hundreds of domains can be built. Many applications in science and
engineering rely on large memory bandwidth; computational fluid
dynamics (CFD) and sparse matrix eigenvalue solvers are typical
examples. However, applications using shared-memory programming models
like, e.g., OpenMP~\cite{openmp}, TBB~\cite{tbb}, or POSIX threads,
should make sure that locality of access is maintained. Massive
performance breakdowns may be observed when nonlocal (inter-LD)
accesses or contention on an LD's memory bus become
bottlenecks~\cite{hpc4se}. One should add that the current OpenMP
standard, although it is the dominant threading model for scientific
user codes, does not contain any features that would enable
ccNUMA access optimizations.

Most operating systems support a \emph{first touch} ccNUMA placement
policy: After allocation (using, e.g., \verb.malloc().), the 
mapping of logical to physical memory addresses is not
established yet; the first write access to an 
allocated memory page will map the page into the locality domain
of the core that executed the write. This makes it straightforward
to optimize parallel memory access in applications that have 
regular memory access patterns. If the loop(s) that initialize
array data are parallelized in exactly the same way and use the
same access patterns as the loops that use the data later,
nonlocal data transfer can be minimized. A prerequisite for
first touch initialization to work reliably is that threads
are not allowed to move freely through the shared-memory machine
but maintain their affinity to the core they were initially 
bound to. Some threading models discourage the use of strong
thread-core affinity, but numerically intensive high-performance 
parallel applications usually benefit from it. Operating systems
often provide libraries and tools to enable a more fine-grained
control over page placement. Under Linux, the \verb.numactl.
command and the \verb.libnuma. library are part of every
standard distribution. 

\begin{table*}[htb]
  \begin{center}
    \begin{tabular}{>{\rule{0pt}{2.3ex}}l|cccc}
 &  & \bfseries Istanbul & \bfseries Nehalem EP & \bfseries Nehalem EX \\
      \hline
      Type                    && Opteron 8431 &  Xeon X5550   & Xeon X7560 \\
      Frequency [\GHZ]        && 2.41   &  2.66   & 2.27 \\
      \hline                 
      Cores per chip          &&   6 &    4  &    8  \\
      Sockets per system      &&   4 &    2  &    4  \\
      L1 size [\KB]            &&  64 &   32  &   32  \\
      L2 size [\KB]            && 512 &  256  &  256  \\
      L3 size [\MB]            &&   5 &    8  &   24  \\
      \hline                   
      L3 cache group [cores]  &&   6 &    4  &    8  \\
      Sockets per system      &&   4 &    2  &    4  \\
      ccNUMA interconnect     &&  HyperTransport (HT) & QuickPath (QPI) & QuickPath (QPI)\\
      \hline                 
	  STREAM copy bandwidth [\GBS]      && & & \\
	  ~~full system		  && 38.6 (NT) &  36.6 (NT) &  33.4  \\
	  ~~socket                &&  9.9 (NT) &  18.9 (NT) &  8.15  \\
    \end{tabular}             
    \caption{Overview of the ccNUMA systems in the test bed. ``NT''
	denotes that nontemporal stores were used in the STREAM
	benchmark as well as for the Jacobi solver test application.}
    \label{tab:systems:overview}
  \end{center}
\end{table*}
\begin{figure}
\centering
\includegraphics*[width=0.8\columnwidth]{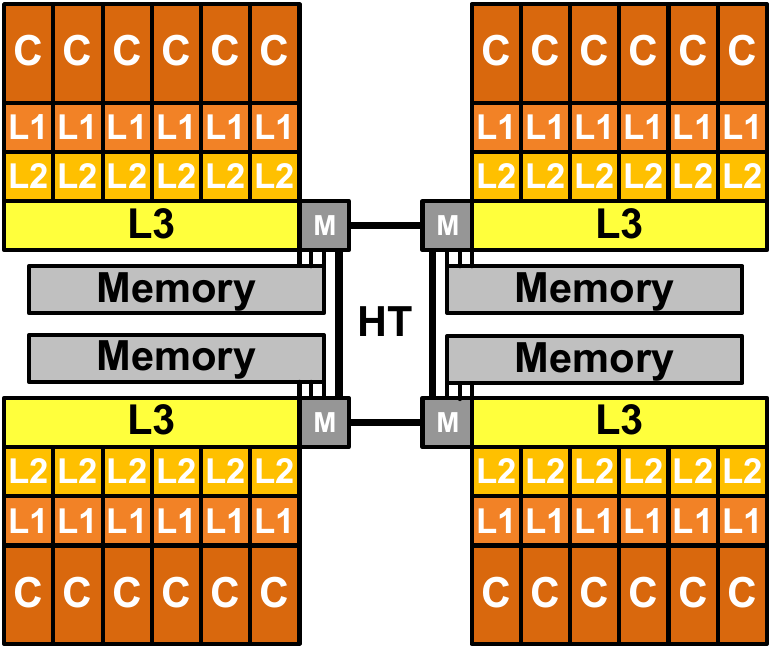}
\caption{\label{fig:nep}Topology of the AMD Istanbul test system with
  four locality domains. The Intel Nehalem
  EX system is very similar but has eight instead of six cores per
  socket, and each socket has direct QPI connections to all other sockets.}
\end{figure}
\begin{figure}
\centering
\includegraphics*[width=0.65\columnwidth]{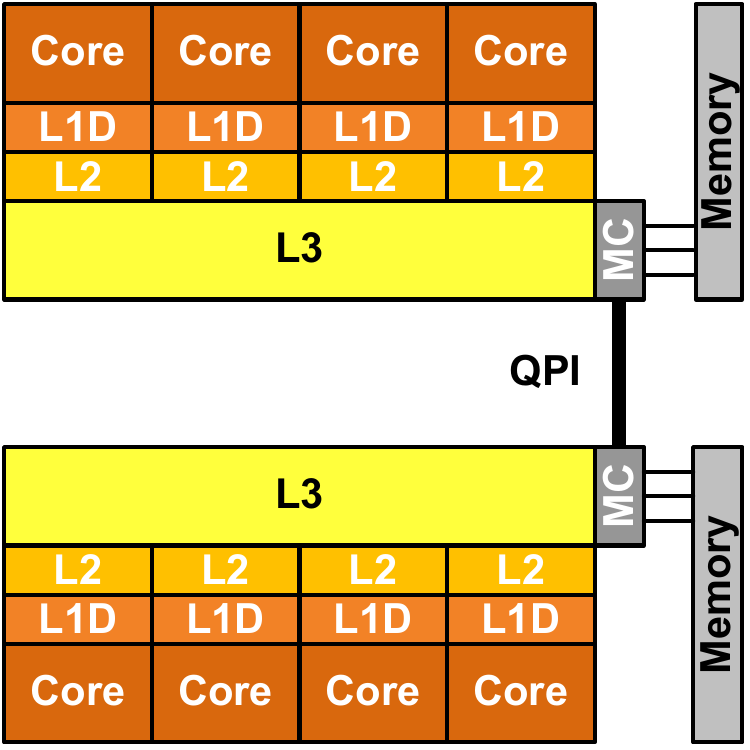}
\caption{\label{fig:i4}Topology of the Intel Nehalem EP test system with two locality
  domains.}
\end{figure}
Unfortunately the ``first touch'' scheme does not work in all cases. Sometimes
memory access cannot be organized in contiguous data streams
or, even if that is possible, the problem itself may be irregular
and show strong load imbalance if a simple static work distribution is
chosen. \emph{Dynamic scheduling} is the general method for 
handling the latter case. The OpenMP standard~\cite{openmp}
provides the \verb.dynamic. and \verb.guided. scheduling types for
worksharing loops, and the \verb.task. construct for task-based
parallelism.  
In Intel Threading Building Blocks (TBB) \cite{tbb}, a
\emph{task} is the central scheduling entity as well, and distribution of
tasks across threads is fully dynamic. 
If the additional overhead for
dynamic scheduling is negligible for the application at hand, these
approaches are ideal on UMA (Uniform Memory Access) systems like the
now outdated single-core multi-socket SMP nodes, or multi-core chips
with ``isotropic'' caches, i.e., where each cache level is either
exclusive to one core or shared among all cores on a chip.
On ccNUMA systems, however, dynamic scheduling leads to nonlocal
memory accesses and contention on the LD's memory buses. The simplest
option to choose is then to distribute memory pages across locality
domains in a cyclic fashion using, e.g., the above-mentioned
NUMA tools, which will lead to at least a certain degree
of parallel memory access. Under Linux, one may write:
\begin{lstlisting}
> env OMP_NUM_THREADS=8 numactl -i 0-3 ./a.out
\end{lstlisting}
This will start the (OpenMP) binary with eight threads and make sure
that memory pages are mapped cyclically across four LDs (0--3).
Initialization inside the program is then insignificant for the 
placement unless special libraries are used, so it may be done
sequentially just as well.

The purpose of this work is to demonstrate that a simple user-level
software layer can make close to optimal ccNUMA page placement
possible even with dynamic scheduling or tasking, by sorting tasks
upon initialization into a number of \emph{locality queues}. We will
show that our scheme works for OpenMP tasking and parallel TBB
constructs, and compare it to the ``affinity partitioner'' in TBB~\cite{tbb},
which has a similar purpose. Contrary to the assumption that tasking
causes ``random'' page access, the order in which tasks are submitted
to the execution thread pool can have a noticeable impact on
performance.

\subsection{Related work}

Using the default first-touch policy with parallel initialization is a
simple optimization technique for memory-bound shared-memory parallel
code, but ccNUMA awareness is unfortunately not yet well established
among application programmers in science and engineering. Moreover,
although introducing multiple execution queues with a work-stealing
scheme on top is not new, the possibilities for enhancing ccNUMA access
locality under dynamic task scheduling \emph{with user code only} and
within the capabilities of current compilers and OS environments have
not been explored in great detail. Most work concentrates on low-level
thread scheduling techniques for various threading models (mostly
OpenMP and Cilk), either runtime-based
\cite{bro_iwomp09,bro_ijpp10,cilknuma}, OS-based \cite{meng2010}, or even
hardware-based~\cite{memtemp}. 
Automatic page migration \cite{yang_ispan09} can enhance locality
significantly, but is again not generally applicable and must
necessarily employ, to varying extent, heuristic methods to decide
about page placement.

The method proposed here consists of a thin software layer that
effectively modifies the task scheduling algorithm employed
by the compiler and runtime system, based on locality information
that can either be supplied by the user or obtained automatically,
depending on the situation.

\subsection{Test bed for performance measurements}

We have chosen three ccNUMA-type systems for performing benchmarks
(see Table~\ref{tab:systems:overview}). The six-core AMD ``Istanbul''
(see Fig.~\ref{fig:i4}) and quad-core Intel ``Nehalem EP'' processors 
(see Fig.~\ref{fig:nep}) have been on the market for
some time; the eight-core ``Nehalem EX'', however, has been introduced
only recently. Our early-access Nehalem EX benchmark system was
equipped with only half the maximum number of memory boards per
socket, which leads to a reduction of the effective main memory
bandwidth by a factor of two. Although of minor importance for
the results presented here, this is of course not a desirable
configuration for a production system.
All systems ran current Linux kernels. The Intel C++ compiler in
version 11.1.064 and TBB version 3.0 (open source variant)
were used for the benchmarks.

All three systems have a similar maximum bandwidth as measured by the
STREAM copy benchmark~\cite{stream}, which models closely the memory
access behavior of the Jacobi solver. Nontemporal stores
(``NT'') were used if appropriate; NT stores bypass the cache
hierarchy and can improve store bandwidth by avoiding the 
write-allocate cache line transfer on store misses.

\subsection{Benchmarking procedure and baseline performance}

\begin{figure*}[tbp]
\centering
\includegraphics*[width=0.95\textwidth]{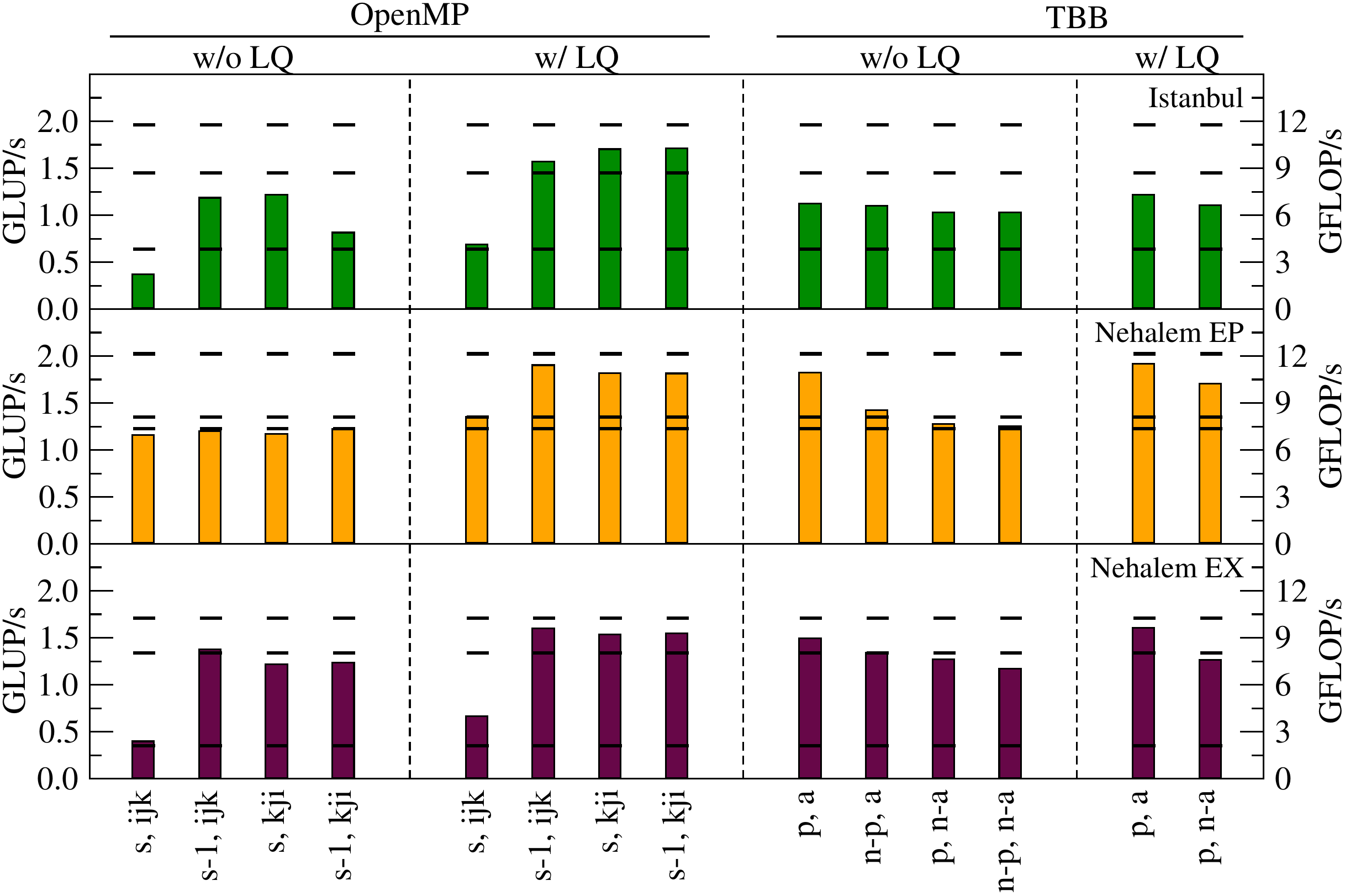}
\caption{\label{fig:perf-all}Performance (median over 100 samples)
  of all code versions
  on the systems in the test bed (all cores utilized). At each
  bar, horizontal lines mark the full-node performance
  under standard OpenMP static worksharing
  with serial initialization in LD0 (bottom), with round-robin page
  placement via \texttt{numactl} (middle), and with correct parallel
  first-touch placement (top). The labels below the columns
  denote static vs. static,1 scheduling for the OpenMP
  initialization loop (``s'' vs. ``s-1''), different task
  submission orders for OpenMP tasking (``ijk'' vs. ``kji''),
  pinned vs. nonpinned TBB threads (``p'' vs. ``n-p''),
  and the use or omission of the affinity partitioner with
  TBB (``a'' vs. ``n-a'').} 
\end{figure*}
As a simple benchmark we choose a 3D six-point Jacobi solver 
with constant coefficients as recently studied extensively
by Datta et al.\ \cite{datta08}\@. 
The site update function,
\begin{eqnarray*}
F_{t+1}(i,j,k) & = & c\cdot\left[F_t(i-1,j,k)+F_t(i+1,j,k)\right. \\
 & &  {}+\left. F_t(i,j-1,k)+F_t(i,j+1,k)\right. \\
 & &  {}+\left. F_t(i,j,k-1)+F_t(i,j,k+1)\right] 
~,
\end{eqnarray*}
is evaluated for each lattice site in a 3D loop nest.
Each site update (in the following called ``LUP'') incurs 
six loads and one store, of which, at large problem sizes,
one load and one store 
cause main memory traffic if suitable spatial blocking is applied.
This leads to a code balance of 8/3 bytes per flop (assuming
that nontemporal stores are used so that a store miss does
not cause a cache line write-allocate transfer), so the code is clearly
memory-bound on all current cache-based architectures.
In what follows we use a problem size of $600^2${}$\times${}$2400$ sites 
($\approx$\,13\,\GB{} of memory for both grids and double precision variables) 
and a blocksize of
600$\times$10$\times$100 ($d_k${}$\times${}$d_j${}$\times${}$d_i$,
with $k$ being the inner [fast] index) 
sites, unless otherwise noted. This is close to the optimal
block dimensions on all architectures considered here.
In a standard OpenMP-parallel
implementation, the update loop nest iterates
over all blocks in turn, and 
standard worksharing parallelization is done over the three
collapsed  blocking loops (first-touch 
initialization is performed via the identical scheme):
\begin{lstlisting}
#pragma omp parallel for \
	collapse(3) schedule(runtime)
  for(int ib=0; ib<no_of_i_blocks; ++ib) {
    for(int jb=0; jb<no_of_j_blocks; ++jb) {
      for(int kb=0; kb<no_of_k_blocks; ++kb) {
        jacobi_sweep_block(ib,jb,kb);
  } } }
\end{lstlisting}
Note that with the standard \verb.k. blocksize being equal to
the extent of the lattice in that direction (which is required
to make best use of the hardware prefetching capabilities
on the processors used), \verb.no_of_k_blocks.
is equal to one.
The \verb.jacobi_sweep_block(). function performs one
Jacobi sweep, i.e., one update per lattice site, over
all sites in the block determined by its parameters.
In case of dynamic loop scheduling
there is a choice as to how parallel first-touch initialization 
should be done; both \verb.static,1. (round robin) and
plain \verb.static. scheduling will be investigated.

Note that this simple benchmark is not a typical application scenario
for tasking, since the load is evenly distributed and parallelization
with standard OpenMP loop worksharing constructs is
straightforward. However, it provides a well-controlled environment
for showing the effects of dynamic scheduling and the limitations of
runtime systems. Moreover, even applications with very regular access
patterns can benefit from task-based parallelism, because functional
decomposition into ``communicating'' and ``computing'' tasks is
greatly simplified. This has been demonstrated recently in the context
of a 3D particle-in-cell code~\cite{alice_cug10}. When using a
threading model together with message passing (MPI) in hybrid
shared/distributed-memory programming it is also vital to reduce
per-node performance variations, since those will limit scalability
of the whole application. We will briefly comment on this problem
below.

For OpenMP we enforced strict thread-core affinity in all benchmark
runs by using the Linux \verb.sched_setaffinity().  function. In
production environments, more user-friendly tools like
hwloc~\cite{hwloc} or likwid-pin~\cite{likwid} are certainly
preferable. In TBB, the concept of a ``thread'' or its affinity
to a piece of hardware is not made explicit for the programmer;
a simple \verb.parallel_for. loop with the number of iterations
equal to the number of spawned threads is repeated until each
thread was assigned a ``dummy'' task for the sole purpose of calling
\verb.sched_setaffinity(). and establishing a fixed thread-core
mapping.

\paragraph{Impact of suboptimal page placement}

The horizontal lines in all panels of Fig.~\ref{fig:perf-all}
illustrate the impact of suboptimal page placement on the solver's
performance. The lowest performance is consistently achieved with
purely sequential initialization, i.e., with a serial initialization
loop, and static worksharing. In this limit, the memory interface of a
single LD becomes a bottleneck and the cores in all but this single
domain have to access their data via the ccNUMA network. Round-robin
placement as established, e.g., with the \verb.numactl.  tool, and
boosts performance significantly by enabling at least some
level of parallelism. Optimal bandwidth utilization is of
course reached with static, parallel first-touch placement, and comes
close to the STREAM copy numbers in
Table~\ref{tab:systems:overview}. On a UMA system (or within a single
ccNUMA domain), all three lines would match. The penalty for
round-robin placement is especially large for the Nehalem EP system,
since it has the strongest ``NUMA effect'' (bandwidth reduction for
nonlocal access). On the other hand, the performance level for
sequential placement is particularly low on Nehalem EX, which can
be attributed to the fact that our EA system is extremely 
bandwidth-starved due to the lack of half the memory boards per LD.
\begin{figure}[tbp]
\centering
\includegraphics*[width=0.95\columnwidth]{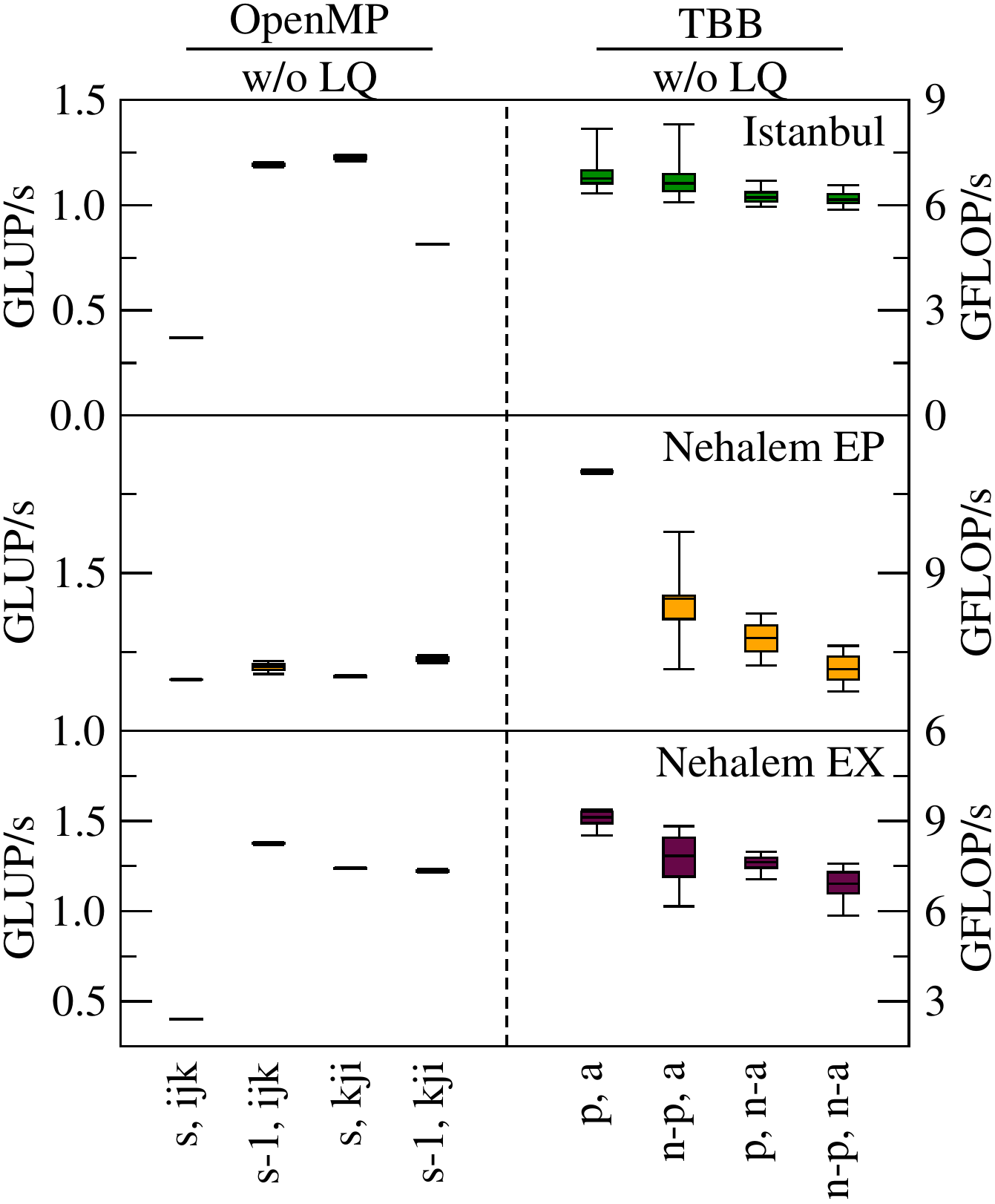}
\caption{\label{fig:nv}Performance variability with OpenMP tasking
	(left) and the TBB \texttt{parallel\_for} construct
	(right). Median, $\pm$25\%, and $\pm$45\% quantiles are
	indicated (100 samples each).}
\end{figure}

Note that the impact of scheduling overhead is not investigated here.
If the amount of work per task is small, dynamic scheduling can
potentially be hazardous for performance~\cite{hpc4se}.

\section{Tasking with OpenMP}

\subsection{Baseline}\label{sec:tasking}

In contrast to standard worksharing loop parallelization, tasking
in OpenMP requires to split the problem into a number of work ``packages'',
called \emph{tasks}, each of which must be submitted to an internal
pool via the \verb.omp task. directive. For the Jacobi solver we define
one task to be a single block of the size specified above. This is
in contrast to standard static loop worksharing, where one parallelized 
loop iteration consisted of several blocks with different coordinates. 

The tasks are produced (``submitted'') by a single thread and consumed 
by all threads and in a 3D loop nest:
\begin{lstlisting}
#pragma omp parallel
{
  #pragma omp single
  {
    for(int ib=0; ib<no_of_i_blocks; ++ib) {
      for(int jb=0; jb<no_of_j_blocks; ++jb) { 
        for(int kb=0; kb<no_of_k_blocks; ++kb) {
          #pragma omp task
            jacobi_sweep_block(ib,jb,kb);
    } } }
  }
}
\end{lstlisting}
Submitting the tasks in parallel is possible but did not make any
difference in the parameter ranges considered here. This parallel
block is actually a ``worksharing'' construct, since all threads that are waiting
in the implicit barrier at the end of the \verb.omp single. construct
execute tasks that have been submitted by the one thread that entered
the \verb.single. region. After finishing the submit loop nest,
this thread will join the others.

In contrast
to the code above, which submits tasks in \verb.jb. direction
first (``ijk''; the single block in \verb.kb. direction
does not count), the loop nest order can be reversed  (``kji''),
leading to a functionally equivalent code. There is also a choice
as to how first-touch initialization should be performed, so we compare
\verb.static. and \verb.static,1. scheduling (``s'' vs. ``s-1'') 
for loop initialization.
The left column of panels in Fig.~\ref{fig:perf-all} shows performance 
results on all platforms. The four combinations of
ijk/kji submit order with static/static,1 initialization are
indicated below the graph. In general, this code is never faster
than standard static worksharing with round-robin placement.
Combining static initialization with ijk submit order seems to be
especially unfortunate.

The large impact of submit and initialization orders can be explained
by assuming that there is only a limited number of ``queued'', i.e.,
unprocessed tasks allowed at any time. In the course of executing the 
submission loop, this limit is reached very quickly and the
submitting thread is used for processing tasks for some time.
From our measurements, the limit is set to roughly 256 tasks with
the compiler used (current GNU compilers have the same limit). 
One \verb.ib.-\verb.jb. layer of the grid comprises
60 tasks (with the chosen problem and block sizes), and 
240 layers are available, which amounts to 14400 tasks in total.
With static scheduling on initialization, one block of 256 consecutive tasks is
usually associated with a single locality domain (rarely two), hence
the serialization of memory access. Choosing \verb.static,1. 
scheduling for initialization, each row of $t$ consecutive blocks
($t$ being the number of threads per socket) 
is placed into a different locality domain, but 256 tasks comprise only
slightly more than four layers. Assuming that the order of execution
for tasks resembles \verb.static,1. loop workshare scheduling 
because each thread is served a task in turn, the number of LDs
to be accessed in parallel is limited (although it is hard to predict
the actual level of parallelism, since it is also influenced 
by the number of threads per LD)\@. Finally, by choosing the
kji submission loop order, consecutive tasks cycle through
locality domains, and parallelism is as expected from dynamic
loop scheduling. In all cases, performance variability is 
surprisingly small (see left panel in Fig.~\ref{fig:nv}).
 
These observations document that it is nontrivial to employ tasking on
ccNUMA systems and reach at least the performance level of standard
dynamic loop scheduling or round-robin page placement. In the next
section we will demonstrate how task scheduling under locality
constraints can be optimized by ``overriding'' part of the OpenMP
task scheduling by user program logic.

\subsection{OpenMP tasking with locality queues}\label{sec:queues}

Each task,
which equals one lattice block (or tile) in our case, is associated
with a C++ object (of type \verb.block.) and equipped
with an integer locality variable. This variable denotes the locality 
domain the block was placed in upon initialization. The submission loop
now takes the following form:
\begin{lstlisting}
#pragma omp parallel
{
  #pragma omp single
  {
    for(int ib=0; ib<no_of_i_blocks; ++ib) {
      for(int jb=0; jb<no_of_j_blocks; ++jb) { 
        for(int kb=0; kb<no_of_k_blocks; ++kb) {
          block *b = blocks[ib][jb][kb];
          queues[b->locality()].enqueue(b);
          #pragma omp task
            process_block_from_queue(queues);
    } } }
  }
}
\end{lstlisting}
The \verb.queues. object is a \verb.std::vector<>. of \verb.std::queue<>.
objects, each associated with one locality domain, and each
protected from concurrent access via an OpenMP lock. Calling the
\verb.enqueue(). method of a queue appends a block object to it. As shown above,
blocks are sorted into those \emph{locality queues} according to their
respective locality variables. One OpenMP task, executed by
the \verb.process_block_from_queue(). function, now consists of two parts:
\begin{enumerate}
\item Figuring out which LD the executing thread belongs to
\item Dequeuing the oldest waiting block in the locality queue belonging 
    to this domain and calling \verb.jacobi_sweep_block(). for it
\end{enumerate}
If the local queue of a thread is empty, other queues are tried
in a spin loop until a block is found (``work stealing''):
\begin{lstlisting}
void process_block_from_queue(locality_queues \
				&queues) {
  // ...
  bool found=false;
  block *b;
  int ld = ld_ID[omp_get_thread_num()];
  while (!found) {
    found = queues[ld].dequeue(p);
    if (!found) {
      ld = (ld + 1) % queues.size();
    }
  }
  jacobi_sweep_block(b->ib, b->jb, b->kb);
}
\end{lstlisting}
The global \verb.ld_ID. vector must be preset with a correct 
mapping of thread numbers to locality domains. 
It is possible with the described scheme that some task executes a
block just queued before the corresponding task is actually
submitted. This is however not a problem because the number of
submitted tasks is always equal to the number of queued blocks, and no
task will ever be left waiting for new blocks forever.

Note that scanning other queues if a thread's local queue is empty
gives load balancing priority over strict access locality, which may or
may not be desirable depending on the application. The team of threads
in one locality domain shares one queue, so scheduling is still purely
dynamic inside an LD.

The second column of panels in Fig.~\ref{fig:perf-all} shows 
performance results:
For \verb.static. initialization and the ijk submission order, 
the limited overall number of waiting tasks has the same consequences as 
with plain tasking (see Sect.~\ref{sec:tasking})\@. In this case,
although the queuing mechanism is in effect, a single queue holds most
of the tasks at any point in time. All threads are served from this queue
and thus mostly execute in a single LD\@. However, using
the alternate kji submission order or \verb.static,1. 
initialization, all queues are fed in parallel and threads can 
always be served tasks from their local queue. Performance then comes
close to static scheduling within a 10\,\% margin.

One should note that a similar effect could have been achieved with
nested parallelism, using one thread per LD in the outer parallel
region and several threads (one per core) in the nested region.
However, we believe our approach to be more powerful and easier 
to apply if properly wrapped into C++ logic that takes care of 
affinity and work distribution. Moreover, the thread pooling strategies
employed by many current compilers inhibit sensible
affinity mechanisms when using nested OpenMP constructs.

\section{Tasking with TBB}

\subsection{Baseline and affinity partitioner}

The universal TBB construct for task-parallel execution
is the \verb.parallel_for. function. Initializing
all blocks by ``first touch'' and performing a domain
sweep looks as follows:
\begin{lstlisting}
tbb::parallel_for(
  tbb::blocked_range_3d<int>(
               0, no_of_i_blocks, 1, 
               0, no_of_j_blocks, 1, 
               0, no_of_k_blocks, 1), 
  touch_block(blocks) );

tbb::parallel_for(
  tbb::blocked_range_3d<int>(
               0, no_of_i_blocks, 1, 
               0, no_of_j_blocks, 1, 
               0, no_of_k_blocks, 1),
  update_block(blocks) );
\end{lstlisting}
The \verb.tbb::blocked_range_3d<>. object encodes the way the
three-dimensional domain (of blocks) is cut into subdomains.
Here we have specified that the smallest unit in each 
coordinate direction is a single block.
In TBB the user must provide a C++ class that implements
\verb.operator(). (i.e., a \emph{functor}), which takes a reference to the range
object and performs the actual ``work'':
\begin{lstlisting}
class update_block
{
  blocks & m_blocks; 

public:
  update_block(blocks & b) 
    : m_blocks(b) {}

  void operator()(tbb::blocked_range_3d<int> 
                                    & subrange) {
    // ... iteration loop nest 
    //     over subrange -> bi, bj, bk
      jacobi_sweep_block(ib, jb, kb);
    // ... end iteration loop nest
  }
// ...
};
\end{lstlisting}
The \verb.subrange. parameter to the functor may encode a single block
or a consecutive range of blocks along all coordinates; this is a
decision made at runtime by TBB.

The third column of panels in Fig.~\ref{fig:perf-all} shows
performance results for TBB with the scheme just described, comparing
the situation with and without binding threads to cores (``p''
vs. ``n-p'') and without using the affinity partitioner
(``n-a'', see below). Since first-touch placement is done
via a \verb.parallel_for. loop, page mapping is dynamic
and performance is close to the round-robin placement
case with standard OpenMP worksharing, as expected. The
mediocre results on the Istanbul system are surprising;
it is as yet unclear why TBB should perform worse than
OpenMP with our locality optimizations employed.

TBB provides a user-friendly way to specify that affinity
information is important for performance. The \verb.tbb::parallel_for.
function takes an optional ``partitioner'' argument, which 
can be set to \verb.tbb::affinity_partitioner.. In this case
TBB stores information about thread-task affinity in an internal
data structure on the first call to \verb.tbb::parallel_for..
On subsequent parallel loops, the scheduler tries to map tasks to 
the same threads as before, thereby establishing 
access locality automatically. The affinity partitioner
must thus be specified on both the initialization and
update loops. The third column of panels shows
performance results with this optimization (``a''),
with and without binding threads to cores (``p''
vs. ``n-p''). Obviously the affinity partitioner can 
significantly improve locality of access and is able to match
the performance of OpenMP tasking with locality queues.

\subsection{TBB tasking with locality queues}

It is possible to adapt the locality queue mechanism to TBB as well,
by letting the \verb.update_block(). functor enqueue the blocks in the
assigned subrange into the appropriate locality queues, and updating
the same number of blocks (preferably) from the executing thread's
local queue. Instead of \verb.std:queue<>., the 
\verb.tbb::concurrent_queue<>. container is used here since it provides
automatic fine-grained locking. However, the performance benefit compared to the
affinity partitioner is marginal (see the fourth column of panels in
Fig.~\ref{fig:perf-all}).  This can be attributed to the fact that
submission order (as defined in the OpenMP tasking versions) cannot be
controlled in this setting. Using a one-dimensional partitioner or a
\verb.parallel_do. construct could enable finer control over page
placement, but the expected additional benefit is small.

\section{Summary and outlook}

We have demonstrated how locality queues can be employed to optimize
parallel memory access on ccNUMA systems when OpenMP tasking or the
TBB \verb.parallel_for. construct is used.
Locality queues substitute the uncontrolled, dynamic task scheduling
by a static and a dynamic part. The latter is mostly restricted to the
cores in one NUMA domain, providing full dynamic load balancing on the
locality domain (LD) level. Scheduling between domains is static, but
load balancing is given priority over strictly local access by a work
stealing scheme. The larger the number of threads per LD, the more
dynamic the task distribution, so our scheme will get more interesting
in view of future many-core processors. Using locality
queues with TBB's \verb.parallel_for. construct does not outperform
the built-in affinity partitioner, but the impact on \verb.parallel_do.
cannot be inferred from this result, and is yet to be investigated.
Note that the concept would in principle work also without 
thread-core affinity because the current locality domain ID of a thread
could be determined at any time, and the static mapping of threads to
LDs would become obsolete.

Future work encompasses the application of the concept to real
application codes, notably sparse matrix eigenvalue solvers, where
load balancing and overlapping computation with communication 
may be achieved in a natural way by tasking.
Further potentials, not restricted to ccNUMA architectures, may be found in 
the possibility to implement temporal blocking (doing more than
one time step on a block to reduce pressure on the memory subsystem
\cite{twh10}) by associating one locality queue to a number of cores
that share a cache level. As an advantage over static temporal blocking,
no frequent global barriers would be required.\vspace*{0.5cm}

\noindent{\bfseries\large Acknowledgments}\vspace*{0.3cm}

\noindent Fruitful discussions with Michael Meier, Gerhard Wellein and Thomas
Zeiser are gratefully acknowledged. We thank Intel Germany
for providing early access hardware and technical support.
This work was supported by BMBF via grant No.\ 01IH08003A (project
SKALB).

%
%

\end{document}